\documentstyle[11pt]{article}
\newcommand{\be}{\begin{equation}}
\newcommand{\ee}{\end{equation}}
\newcommand{\bea}{\begin{eqnarray}}
\newcommand{\eea}{\end{eqnarray}}
\newcommand{\beaa}{\begin{eqnarray*}}
\newcommand{\eeaa}{\end{eqnarray*}}
\newcommand{\ol}{\overline}
\newcommand{\ul}{\underline}
\newcommand{\al}{\alpha}
\newcommand{\bt}{\beta}
\newcommand{\la}{\lambda}

\newcommand{\Ga}{\Gamma}
\newcommand{\si}{\sigma}
\newcommand{\om}{\omega}
\newcommand{\fo}{\frac{\om}2}

\newcommand{\lag}{\langle}
\newcommand{\rag}{\rangle}
\newcommand{\bl}{\biggl(}
\newcommand{\br}{\biggr)}
\newcommand{\cbl}{\biggl\{ }
\newcommand{\cbr}{\biggr\} }
\newcommand{\cR}{\check{R}}
\newcommand{\cP}{\check{P}}
\newcommand{\tm}{\tilde{\mu}}
\newcommand{\nn}{\nonumber}
\newcommand{\fns}{\footnotesize}
\newcommand{\scs}{\scriptsize}
\newcommand{\shs}{\shortstack}

\newcommand{\rank}{{\hbox{rank}}}
\newcommand{\q}{{{}_q}}

\newcommand{\qi}{{{}_{q^\imath}}}
\newcommand{\qtwo}{{{}_{q^2}}}
\newcommand{\qthree}{{{}_{q^3}}}
\advance\textheight by \topskip
\makeatletter

\@addtoreset{equation}{section}
\def\section{\@startsection {section}{1}{\z@}{-3.5ex plus -1ex minus
 -.2ex}{2.3ex plus .2ex}{\large\bf\centering}}
\def\subsection{\@startsection{subsection}{2}{\z@}{-3.25ex plus%
 -1ex minus -.2ex}{1.5ex plus .2ex}{\bf}}
\def\subsubsection{\@startsection{subsubsection}{3}{\z@}{-3.25ex plus%
 -1ex minus -.2ex}{1.5ex plus .2ex}{\sl}}
\makeatother

\begin{document}
\def\ss{\scriptstyle}
\baselineskip 17pt
\parindent 12pt
\parskip 10pt

{
\parskip 0pt
\newpage
\begin{titlepage}
\begin{flushright}
DAMTP-95-49\\
CRM-2314\\
hep-th/9509007\\
revised November 95\\[3mm]
\end{flushright}
\vspace{.4cm}
\begin{center}
{\Large {\bf
Twisted  algebra  $R$-matrices  and  $S$-matrices for  $b_n^{(1)}$
\vspace{.2cm}\\
affine Toda  solitons  and  their  bound  states
}}\\
\vspace{1.2cm}
{\large G.M. Gandenberger$^{\dagger}$}\footnote{e-mail:
G.M.Gandenberger@damtp.cam.ac.uk}\\
\vspace{2mm}
{\large N.J. MacKay$^{\dagger,*}$}\footnote{e-mail:
N.J.MacKay@damtp.cam.ac.uk}\\
\vspace{2mm}
{and}\\
\vspace{2mm}
{\large G.M.T.~Watts$^{\dagger,*}$}\footnote{e-mail:
G.M.T.Watts@damtp.cam.ac.uk}\\
\vspace{3mm}
{\em $\dagger$ Department of Applied Mathematics and Theoretical Physics,}\\
{\em Cambridge University}\\
{\em Silver Street, Cambridge, CB3 9EW, U.K.}\\
\vspace{5mm}
{\em $*$ Centre de Recherches Math\'ematiques,
Universit\'e de Montr\'eal,}\\
{\em C.P. 6128 Succ. Centre Ville, Montr\'eal (Qu\'ebec) H3C 3J7, CANADA.}\\
\vspace{1.5cm}
{\bf{ABSTRACT}}
\end{center}
\begin{quote}
We construct new $U_q(a^{(2)}_{2n-1})$ and $U_q(e^{(2)}_6)$ invariant
$R$-matrices and comment on  the  general  construction  of
$R$-matrices  for  twisted  algebras. We use the former  to  construct
$S$-matrices  for  $b^{(1)}_n$ affine  Toda  solitons  and  their  bound
states, identifying the lowest breathers with the $b^{(1)}_n$ particles.
\end{quote}
\vfill
\end{titlepage}
}

\section{Introduction}
\label{sec:one}
\setcounter{footnote}{0}

Affine Toda field theories (ATFTs) based on affine algebras $\hat g$
have quantized $\hat g^\vee$ charge algebras\footnote{The dual
${\cal A}^\vee$ of an algebra ${\cal A}$ is obtained by replacing
its simple roots by co-roots, equivalent to reversing the arrows on its
Dynkin diagram.}
\cite{charge}, and consequently $S$-matrices for affine Toda solitons can be
constructed from $U_q(\hat g^\vee)$ \mbox{$R$-matrices}, as has been done for
$a_n^{(1)}$ by Hollowood~\cite{hollo93}.
For more details we refer the reader to \cite{gande95b,corri94}.

For cases where $\hat g$ is untwisted and nonsimply-laced, the least
understood of the ATFTs, twisted algebra $R$-matrices must be used.
It has been a longstanding problem, recently solved by Delius, Gould
and Zhang \cite{deliu95b}, to construct $R$-matrices for quantized twisted
affine algebras. In this paper we build on their results
to produce new $R$-matrices associated with the quantum deformations
of the algebras $a_{2n-1}^{(2)}$ and $e_6^{(2)}$.

In particular our $a_{2n-1}^{(2)}$ $R$-matrices allow us
to construct $S$-matrices for $b_n^{(1)}$ solitons and their bound states.
Our main result here is that the lowest scalar bound states (`breathers')
may be identified with the $b_n^{(1)}$ ATFT quantum particles.

The layout of the paper is as follows:
in section \ref{sec:oneA} we give our conventions for the quantized
universal enveloping algebras.
In section \ref{sec:two} we
summarize the previous constructions of $R$-matrices for untwisted and
twisted algebras and in section \ref{sec:thr} give our new
constructions. We finish this section with some general comments on
the construction of $R$-matrices.
After some more general material in section 5, in section
\ref{sec:fou} we give details of the $b_n^{(1)}$ soliton
and breather $S$-matrices constructed using the $a_{2n-1}^{(2)}$
$R$-matrices found in section \ref{sec:thr}, and we finish this
section with some comments on the status of imaginary coupling affine
Toda field theories in general.

We follow the conventions of \cite{gande95b}, in which other details such
as Coxeter and dual Coxeter numbers may be found.

\section{Quantized enveloping algebras}
\label{sec:oneA}

For ease of notation we shall everywhere write $\q{\cal A}$ for
$U_q({\cal A})$.

The generators of $\q g$ are $e_i,f_i,h_i$,
$i=1,..,\rank(g)$,
satisfying the $q$--deformed Serre relations (see \cite{deliu95b}
for these and more details) and
\be
  [h_i,e_j]=(\al_i,\al_j)e_j
\;,\;\;
  [h_i,f_j]=-(\al_i,\al_j)f_j
\;,\;\;
  [h_i,h_j]=0
\;,\;\;
\label{eq:def1}
\ee
\be
  [e_i,f_j]=\delta_{ij}\frac{q^{h_i}-q^{-h_i}}{q_i-q_i^{-1}}
\;,\;\;
  q_i=q^{\frac 12(\al_i,\al_i)}
\;,
\label{eq:def2}
\ee
with coproduct%
\footnote{Note that this is the opposite coproduct to that used in
\cite{charge}, so that the $q$ in that paper is the inverse of $q$
here}
\bea
 \Delta(e_i)
= q^{-h_i/2}\otimes e_i+ e_i\otimes q^{h_i/2}
\;,&&
  \Delta(f_i)
= q^{-h_i/2}\otimes f_i+ f_i\otimes q^{h_i/2}
\;,\;\; \nn\\
    \Delta(q^{\pm h_i/2})
&\!\!=\!\!&
    q^{\pm h_i/2}\otimes q^{\pm h_i/2}
\;.
\label{eq:def3}
\eea
The generators of $\q g^{(k)}$ for an affine algebra also satisfy
equations (\ref{eq:def1},\ref{eq:def2},\ref{eq:def3}), but where the
index $i$ now runs from 0.
Since the roots appear explicitly in (\ref{eq:def1}), we have to
choose a normalization, which is that the long roots of a finite
algebra $g$ satisfy $|\alpha_{long}|^2 = 2$, whereas for those of
an affine algebra $g^{(k)}$ satisfy $|\alpha_{long}|^2 = 2k$.

We shall also need Reshetikhin's result \cite{Resh1} on the spectral
decomposition of the $\q g$ $R$-matrix. This holds in any tensor
product of two irreducible representations $V_\lambda \otimes V_\mu =
\oplus_\nu V_\nu$ for which each irreducible representation $V_\nu$
occurs with multiplicity at most one, in which case we have
\be
  R_{\lambda\mu}
= \sum_{\nu} \epsilon(\nu) q^{C_2(\nu) - C_2(\lambda) - C_2(\mu)}
\;,
\ee
where $\epsilon(\nu)=\pm1$, and where
$C_2(\lambda)$ is the Casimir operator normalized so that
$C_2(\hbox{adjoint})=h^\vee$, the dual Coxeter number of $g$.

\section{$R$-matrices for untwisted algebras}
\label{sec:two}

The construction of Delius et al.~\cite{deliu95b} is an extension of the
Tensor Product Graph method \cite{zhang91,macka91,deliu94}.
We outline this method in the case of untwisted algebras now.

Let $g^{(1)}$ be an untwisted affine algebra. It is clear from the
defining generators and relations that the generators
$e_{i},f_{i},h_{i}$ with \hbox{$i\!=\!1,..,\rank(g)$}
generate a $\q g$ subalgebra of $\q g^{(1)}$.

The physical particles transform as finite-dimensional
$\q g^{(1)}$ representations, and so the $R$-matrices of
interest are those in the tensor product $V_a\otimes V_b$ of
finite-dimensional $\q g^{(1)}$ modules.

Given a representation $\pi_a$ of $\q g^{(1)}$ on a vector space
$V_a$, we can then find another representation $\pi_a(x)$ by taking
\[
  \pi_a(x)( e_0 ) = x      \pi_a ( e_0 )
\;,\;\;
  \pi_a(x)( f_0 ) = x^{-1} \pi_a ( f_0 )
\;,\;\;
  \pi_a(x)( h_0 ) =        \pi_a ( h_0 )
\;,
\]
\be
  \pi_a(x)( \q g ) = \pi_a ( \q g )
\;,\;\;
\ee
which is simply a gauge transformation. $x$ is the
spectral parameter.

We take the $\cR_{a,b}(x)$ to satisfy
\be
  \cR_{a,b}(x) ( \pi_a(x) \otimes \pi_b(1) )
  \left(\Delta( \q g ) \right)
= ( \pi_b(1) \otimes \pi_a(x) )
  \left(\Delta( \q g ) \right) \cR_{a,b}(x)
\;,\;\;
\ee
\be
  \cR_{a,b}(x) \cR_{ba}(x^{-1}) = {\bf 1}
\;,\;\;
  \cR_{a,b}(0) = \cR_{a,b}
\;,
\ee
where $\cR_{a,b}$ is the $R$-matrix of $\q g$.
This fixes $\cR_{a,b}(x)$ up to an overall factor $f(x)$ such that
$f(x) f(x^{-1}) =1, f(0)=1$.
Since {\bf P}$\,\cR_{a,b}$ commutes with the action of
$\q g$ (where {\bf P} is the permutation matrix),
{\bf P}$\cR_{a,b}(x)$ must be a sum of projectors onto the irreducible
$\q g$-components of $V_a \otimes V_b$.
We denote the $\q g$--projector onto representation $V_{c}$ by
{\bf P}$\cP_c$,
so that
\be
  \cR_{a,b}(x) = \sum_{c_i} f_i(x) \cP_{c_i}
\;.
\ee

If each irreducible $\q g$-representation $V_c$ occurs no more than
once in $V_a \otimes V_b = \sum_i V_{c_i}$, then we say that it is
multiplicity-free. It is only in this case that we can determine
$\cR_{a,b}$ by the Tensor Product Graph method alone. An immediate
requirement for the construction of $R_{a,a}$ is that $V_a$ is itself
irreducible with respect to $\q g$.
If $V_a\otimes V_b$ is multiplicity free then we find%
\footnote{%
n.b. this is different from the results in \cite{deliu95b} as we have
chosen to normalize $C_2$ differently here.
}
\be
  \frac{ f_j (x) }{ f_i (x) }
=
  \left\lag C_2( V_{c_i} ) - C_2( V_{c_j} )\right\rag_{
{\epsilon(c_i)\epsilon(c_j)}}
\;,
\label{eq:fratio}
\ee
for each pair of representations $V_{c_i}$ and $V_{c_j}$ such that
\hbox{$V_{c_j} \subset V_{c_i} \otimes \tilde V$}, where  $\tilde V$
is the $\q g$ representation corresponding to the extra
generator $e_0$, and where
\be
  \lag a \rag_\pm
= \frac{ 1 \pm x q^a }{x \pm q^a }
\;.
\label{eq:bk}
\ee
The $R$-matrix is then determined up to an overall scalar factor,
which is usually fixed by taking one of the modules to have coefficient $1$.

A standard way of presenting this information is the Tensor Product
Graph itself (or TPG for short), in which the components $V_{c_i}$ and
$V_{c_j}$ are connected by an arrow from $i$ to $j$ if and only if
\hbox{$V_{c_j} \subset V_{c_i} \otimes  \tilde V$}, and the arrow is
labelled by $a$.
(The TPG edge is thus directed, and in the case that $\tilde V$ is
self-conjugate, as will always be true for us, an edge with label $a$
is equivalent to the reverse-directed edge with label $-a$).  We choose
the module with coefficient $1$ to be that with the highest weight,
and write the TPG with arrows directed only away from it.
In \cite{deliu94} it was argued that in the untwisted case the
product $\epsilon(c_i)\epsilon(c_j)$ for two components joined by a
TPG edge is always $-1$.

\subsection{$R$--matrices for twisted algebras}

A twisted quantum affine Lie algebra $\q g^{(k)}$ has generators
$e_i,f_i,h_i$ for $i=0,..,l$ also satisfying
equations (\ref{eq:def1}), (\ref{eq:def2}), (\ref{eq:def3}), but where
$\al_i$ are now the simple roots of a twisted affine algebra.

Exactly as in the case of an untwisted quantum affine Lie algebra,
there are subalgebras of $\q g^{(k)}$ generated by
$e_{i'},f_{i'},h_{i'}$ where $\{i'\}$ are any subset of
$\{0,..,l\}$.
Such a subalgebra is a direct sum
\be
 \oplus  \;\left(\; {}_{q_i} h_i \;\right)
\;,
\ee
where the Dynkin diagrams of $h_i$ are the different components
of the subdiagram of that of $g^{(k)}$ consisting of the nodes
$\{i'\}$.
There is one subtlety involved%
\footnote{This was passed over in \cite{deliu95b}}%
, which is that the roots $\{\al_{i'}\}$
of any particular subalgebra ${}_{q_i}h_i \subset \q g^{(k)}$  may not
agree with the specific normalization for the root lengths we have
chosen. If $2 \imath$ is the square of the longest root in the set
$\{\al_{i'}\}$, then we have
\be
q_i = q^\imath
\;.
\ee
This can be different for the different components $h_i$ in $g^{(k)}$.

In particular, as noted by Delius et al. in \cite{deliu95b}, one can
always obtain $g_0$ as a subalgebra of $g^{(k)}$, where
$g_0$ is the subalgebra of $g$ invariant under the order $k$
automorphism defining the twist,
\be
  {}_{q^\imath}g_0 \subset \q g^{(k)}
\;.
\ee

The TPG method now carries on as before, but the
representation $\tilde V$ in which the extra generator of the twisted
quantum affine algebra transforms is no longer the adjoint
representation of $\qi g_0$. If a $\q g_0^{(1)}$ $R$-matrix exists
for the same $g_0$-modules its TPG will therefore be
topologically different: the two TPGs will correspond to different
`Baxterizations' \cite{ge90} of the centralizer algebra (the
Birman-Wenzl-Murakami algebra in the $b$-, $c$- and $d_n$ cases).

The crux of the argument of Delius et al.~\cite{deliu95b},
was to establish the relative parities of the components of
the TPG in this case.
(Note that in their notation Lie algebras are denoted by $L$ in
contrast to our $g$.)
They observed that, when the $\q g^{(k)}$ representations are
irreducible $\qi g_0$ representations, the relative parity of two linked
components $V_{c_i}$ and $V_{c_j}$ in the TPG is $+1$ if they are part
of the same $\q g$ representations, and
$-1$ if they are parts of different $\q g$ representations. Following
this observation, the rest of the argument is identical to that in
the case of untwisted algebras, with the understanding that the
`block' in equation (\ref{eq:fratio})
should be replaced by
\be
  \frac{ f_j (x) }{ f_i (x) }
=
  \left\lag\; \imath
  \;\left(\; C_2( V_{c_i} ) - C_2( V_{c_j} )\;\right)\;
  \;\right\rag_\pm
\;,
\label{eq:fratio2}
\ee
for each pair of representations $V_{c_i}$ and $V_{c_j}$ such that
\hbox{$V_{c_j} \subset V_{c_i} \otimes \tilde V$}, where  $\tilde V$
is the $\qi g_0$ representation corresponding to the extra
generator $e_0$.

Our extension of the work of \cite{deliu95b} is based on
the observation that there may be more than one choice of finite
dimensional subalgebra $h$ of $g^{(k)}$ such that
$\rank(h) = \rank(g_0)$: such an $h$ is obtained by deleting one
node from the Dynkin diagram of $g^{(k)}$.
In this paper we only consider the cases where the $h$ we obtain in
this way is simple;
we give these, along with the index
$\imath$ in each case, in the table \ref{tab:mults} below: $g_0$ is
the choice given in \cite{deliu95b}, and $g_0'$ is the only simple
alternative. We recall that $\q  g^{(1)}$-modules may be reducible as
$\q g$-modules%
\footnote{For a full listing of those currently known, see
\cite{chari91}.},
and that, when the $R$-matrices are used to construct $S$-matrices,
it is the fundamental $\q g^{(1)}$-modules (those whose highest
component is a fundamental $\q g$-module) which correspond to the
physical particle multiplets. We observe that the twisted algebra
multiplets, the $\q g^{(k)}$-modules, are always a subset of the
$\q g^{(1)}$-modules, and in table \ref{tab:mults} we give the
decomposition of these as $\q g$-, $\qi g_0$- and
$\qi g_0'$-modules, where $V_\mu$ is the module with highest
weight $\mu$, and the $\lambda_i$ are the fundamental weights. For
ease of notation we also write $V_i \equiv V_{\lambda_i}$ ; $V_0$ is
the singlet.

{
\renewcommand{\arraystretch}{1.3}
\headsep -0.5in
\newcommand{\Mults}{\hbox{\small Multiplets}}
\begin{table}[hbt]
\caption{Subalgebras and multiplets}
\label{tab:mults}

\[
\begin{array}{|c|l|l|l|}
\hline
  g^{(k)}      & g        & g_0   & g_0' \\
\hline\hline

a_{2n}^{(2)}   & a_{2n}   & b_n   & c_n  \\
\imath         &          &  1    &  2   \\
\tilde V       &          & V_{2 \lambda_1} & V_1 \\
\hline
\Mults     & V_r\;,\;\; r=1,..,\,n\!\!-\!\!1
               & V_r\;,\;\; r=1,..,\,n\!\!-\!\!1
               & \oplus_{i=1}^r V_i\;,\;\; r=1,..,\,n\!\!-\!\!1 \\
               & V_n &  V_{2\lambda_n} & \oplus_{i=1}^n V_i \\
\hline\hline

a_{2n-1}^{(2)} & a_{2n-1} & c_n   & d_n \\
\imath         &          &  2    &  1  \\
\tilde V       &          & V_2   & V_{2 \lambda_1} \\
\hline
\Mults     & { V_r\;,\;\; r=1,..,\,n\!\!-\!\!2}
               & \oplus_{i=0}^{[r/2]} V_{r-2i}\;,\;\; r=1,..,\,n\!\!-\!\!2
               &  V_r \;,\;\; r=1,..,\,n\!\!-\!\!2 \\
               & V_{n-1}
               & \oplus_{i=0}^{[(n-1)/2]} V_{n-1-2i}
               & V_{\lambda_{n-1}+\lambda_n} \\
               & V_n & \oplus_{i=0}^{[n/2]} V_{n-2i}
               & V_{2\lambda_{n-1}} \oplus V_{2\lambda_n}   \\
\hline\hline

d_{n+1}^{(2)}  & d_{n+1}  & b_n   & \hbox{---} \\
\imath         &          &  2    &  \\
\tilde V       &          & V_1   & \\
\hline
\Mults     & V_r \oplus V_{r-2}\oplus..\,,\; r=1,..,\,n\!\!-\!\!1
               & \oplus_{i=1}^r V_i\;,\;\; r=1,..,\,n\!\!-\!\!1 & \\
               & V_n & V_n & \\
\hline\hline

e_{6}^{(2)}    & e_6      & f_4      & c_4 \\
\imath         &          &  2       &  2  \\
\tilde V       &          & V_4      & V_4 \\
\hline
\Mults     & V_1      & V_4 \oplus V_0 & V_2 \\
               & V_2 \oplus V_0 & V_1\oplus V_4\oplus V_0
               & V_4 \oplus V_{2\lambda_1}\oplus V_0 \\
               & V_3\oplus V_6 & V_3\oplus V_1\oplus 2V_4\oplus V_0
               & V_{\lambda_1\oplus\lambda_3}\oplus V_{2\lambda_1}\oplus V_2\\
               & V_4\oplus V_{\lambda_1+\lambda_6}\oplus 2V_2\oplus V_0
               & \cdots
               & \cdots
\\
\hline\hline

d_{4}^{(3)}    & d_4      & g_2      & a_2 \\
\imath         &          &  3       &  1  \\
\tilde V       &          & V_2      & V_{3\lambda_1} \\
\hline
\Mults     & V_1      & { V_2\oplus V_0 } & 8 \\
               & V_2\oplus V_0 & V_1 \oplus 2 V_2 \oplus V_0
               & { 10 \oplus \bar{10} \oplus 8 \oplus 1 }\\
\hline\end{array}
\]
In this table and throughout the paper we used \cite{Lie} and
\cite{Patera} for Lie algebra information. \\
The Dynkin diagrams and
their numberings are given in table 1 of \cite{macka94}.
\end{table}
}

\section{$R-$matrices from $g_0'$}
\label{sec:thr}

We now consider, for each of the twisted algebras in turn,
whether using $g_0'$ instead of $g_0$ enables us to
construct new $R-$matrices.
As Delius et al.~point out, we can only hope to construct $R$-matrices
for those $\q g^{(k)}$ multiplets which are $\qi h$-irreducible.

\subsection{$a_{2n-1}^{(2)}$ }

We see that this algebra has a $\q d_n$ as well as a $\qtwo c_n$ subalgebra
and that the multiplets are (mostly) irreducible modules of the
former. Hence we can now construct the TPG for $R_{a,b}$ if $a+b \leq n$,
as in this case $V_a \otimes V_b$ is multiplicity-free. We find that
the TPG is of the now-familiar form first seen in \cite{macka91},

\be
\begin{array}{ccccccccc}
\mu_a+\mu_b          & \to    &
\mu_{a+1}+\mu_{b-1}  & \cdots & \cdots & \to &
\mu_{a+b-1}+\mu_1    & \to    & \mu_{a+b}      \\[2.5mm]
\downarrow           &        &
\downarrow           &        &        &     &
\downarrow           &        &                \\[2,5mm]
\mu_{a-1}+\mu_{b-1}  & \to    &
\mu_a+\mu_{b-2}      & \cdots & \cdots & \to &
\mu_{a+b-2}          &        &                \\[2.5mm]
\vdots               &        &
\vdots               &        &        &     & & & \\[2.5mm]
\downarrow           &        &
\downarrow           &        &        &     & & & \\[2.5mm]
\mu_{a-b+1}+\mu_{1}  & \to    &
\mu_{a-b+2}          &        &        &     & & & \\[2.5mm]
\downarrow           &        &  & &   &     & & & \\[2.5mm]
\mu_{a-b}            &        &  & &   &     & & &
\end{array}
\label{eq:TPG}
\ee
where the $\mu_i = \lambda_i$ (for $i = 1,..,n\!-\!2) $,
$\mu_{n-1} = \lambda_{n-1} + \lambda_n$ ,
and where the node $\mu$ denotes the irreducible representation
$V_\mu$ in all cases except $\mu_n$, which denotes
$V_{2 \lambda_{n-1}} \oplus  V_{2 \lambda_n}$.
The columns of the graph alternate in parity, with the first column
having positive parity.
We obtain
\be
  \cR_{a,b}^{(TPG)}(x)
= \sum_{p=0}^b \sum_{r=0}^{b-p}
  \prod_{i=1}^p \lag a-b+2i \rag_-
  \prod_{j=1}^r \lag 2n-a-b+2j \rag_+
  \cP_{\mu_{a+p-r}+\mu_{b-p-r}}
\;.
\label{specdecom}
\ee

{
\renewcommand{\arraystretch}{1.2}
\headsep -0.5in
\begin{table}[hbt]
\caption{}\label{tab:R11}

\[
\begin{array}{ccc}
\qtwo c_n \subset \qtwo c_n^{(1)} & & \q d_n \subset \q d_n^{(1)}\\[3mm]
{
 \begin{array}{ccc}
V_{2\la_1}^+  & \rightarrow^2 & \hspace{2mm}V_2^- \\[2mm]
\hspace{4mm}\downarrow_{2n+2} && \\[2mm]
V_0^- &&
\end{array} }
& &
{ \begin{array}{ccc}
  V_{2\la_1}^+  &\to^2 & V_2^- \\[2mm]
                &      &\hspace{4mm}\downarrow_{2n-2} \\[2mm]
                &      & V_0^+
\end{array} }
\\[5mm]
&
\begin{picture}(70,70)

\put(0,70){\vector(1,-1){70}}
\put(65,15){\small $q^{2n+2}\mapsto -q^{2n}$}

\put(70,70){\vector(-1,-1){70}}
\put(-55,15){\small $q^{2n-2}\mapsto -q^{2n}$}
\end{picture}

 &
\\[5mm]

\qtwo c_n \subset \q a_{2n-1}^{(2)} & &\q d_n \subset \q
a_{2n-1}^{(2)} \\[3mm]
{ \begin{array}{ccc}
  V_{2\la_1}^+  &\to^2 & V_2^- \\[2mm]
                &      &\hspace{1mm}\downarrow_{2n} \\[2mm]
                &      & V_0^-
\end{array} }
& &
{ \begin{array}{ccc}
  V_{2\la_1}^+               &\to^2 & V_2^- \\[2mm]
  \hspace{1mm}\downarrow_{2n}&      &\\[2mm]
  V_0^+                      &      &
  \end{array} }
\\[8mm]
\end{array}
\]
{\centerline{Relative parities of the different $V$ are indicated
by $V^\pm$.}}
\end{table}
}

As an example, consider $\cR_{1,1}$, which can be constructed using
either the $\qtwo c_n$ or the $\q d_n$ invariance. The relevant
TPGs are given in the second row of table \ref{tab:R11}.
Remembering that for $c_n$,
\be
  \dim V_{2\lambda_1} = n(2n+1)
\;,\;\;
  \dim V_{2} = (n-1)(2n+1)
\;,\;\;
  \dim V_{0} = 1
\;,
\ee
and that for $d_n$,
\be
  \dim V_{2\lambda_1} = (n+1)(2n-1)
\;,\;\;
  \dim V_{2} = n(2n-1)
\;,\;\;
  \dim V_{0} = 1
\;,
\ee
we see that the two $R$-matrices have the same rank for all $x$.
Since these are both $R$-matrices for $\q a_{2n-1}^{(2)}$, they are in
fact similar, being related by a gauge transformation.

Now examine the first row of table \ref{tab:R11}.
The $\qtwo c_n^{(1)}$ $R$--matrix is also $\qtwo c_n$ invariant, but
has a different TPG from the $\qtwo c_n \subset\q a_{2n-1}^{(2)}$
$R$--matrix.
Comparing these two $R$-matrices, we see that they give the
same braid operator in the $x\rightarrow 0$ limit: as noted,
they correspond
to two different `Baxterizations' of this braid operator.

The TPGs of the $\qtwo c_n \subset \qtwo c_n^{(1)}$ (top left in table
\ref{tab:R11}) and $\q d_n \subset \q a_{2n-1}^{(2)}$ (bottom right)
$R$-matrices are superficially similar, but of course the parities
are different and the nodes of the former are $\qtwo c_n$, not $\q
d_n$, modules. We see, however, that if we replace $q^{2n+2}$ by
$-q^{2n}$ in the $x$-dependent coefficients of the former, we obtain
those of the latter.
Furthermore, making this change in the $\qtwo c_n$ BWM algebra
\cite{macka91b} maps it (and thus the projectors onto the various
representations) to the $\q d_n$ BWM algebra. Combining these,
we find that applying $q^{2n+2}\mapsto -q^{2n}$ to the full
$R$-matrix maps the one into the other. We expect this to apply to the
higher $R$-matrices too, a fact which may have significance for affine
Toda solitons.

Similarly, applying $q^{2n-2}\mapsto -q^{2n}$ to the $\q d_n \subset
\q d_n^{(1)}$ $R$-matrix gives us precisely the $\qtwo c_n
\subset \q a_{2n-1}^{(2)}$ $R$-matrix; Jimbo obtains his
 $\q a_{2n-1}^{(2)}$ $R$-matrix \cite{jimbo86} in this way.
This is in contradiction
to the note in \cite{artz95}, which suggested that the arrows
in table \ref{tab:R11} should be vertical, with $q^{2n+2}\mapsto
-q^{2n}$ mapping the $\qtwo c_n \subset \qtwo c_n^{(1)}$
to the $\qtwo c_n \subset \q a_{2n-1}^{(2)}$ $R$-matrix, and so on.

\subsection{$a_{2n}^{(2)}$ }

For $\q a_{2n}^{(2)}$ Delius et al.\ were able to find all the
$R$-matrices since the relevant $\q a_{2n}$ representations are all
irreducible with $g_0 = b_n$
(so that $g_0=so(N+1)$ for $g^{(k)}=a_N^{(2)}$).
They found $R$-matrices which correspond to the TPG (\ref{eq:TPG})
above for {\em all} $a,b$, with $\mu_i=\la_i$ for $i=1,..,n\!-\!1$,
$\mu_n=2\la_n$ and $\mu_i=\mu_{2n+1-i}$ for $i=n\!+\!1,..,2n$. These
$R$-matrices thus include the $a,b\rightarrow 2n\!+\!1\!-\!a\!-\!b$
fusion which does not occur for non-self-dual algebras.

\subsection{$d_{n+1}^{(2)}$ }

There is only one possible simple Lie algebra Dynkin diagram
obtainable by deleting one node from the $d_{n+1}^{(2)}$ diagram,
and so we cannot add to the results of \cite{deliu95b} here, except to
note that with our normalisations,
$\qtwo b_n \subset \q d_{n+1}^{(2)}$,
rather than $\q b_n \subset \q d_{n+1}^{(2)}$ as in \cite{deliu95b}.

\subsection{ $e_6^{(2)}$ }

{}From table \ref{tab:mults} we see that the $27$ dimensional
representation of $\q L=\q e_6$ is an irreducible representation of
$\qi q_0'=\qtwo c_4$, and since the tensor product of two $27$
dimensional $\qtwo c_4$ representations is multiplicity free, we can
construct the TPG for $R_{\ul{27},\ul{27}}$,
\[
\begin{array}{ccccc}
V_{2\lambda_2}       & \to^1  & V_{\lambda_1+\lambda_2} & \to^4 &
                                                        V_2 \\[2,5mm]
\downarrow_3         &        & \downarrow_3            & & \\[2,5mm]
V_4                  & \to^1  & V_{2\lambda_1}          & & \\[2,5mm]
\downarrow_6         &        &                         & & \\[2,5mm]
V_0                  & & & &
\end{array}
\]
where the nodes denote irreducible $\qtwo c_4$ representations. Again
the columns of the graph alternate in parity, with the first column
having positive parity.

\subsection{$d_{4}^{(3)}$ }

In this case we find that the $8$ dimensional representation of
$\q d_4$, which was the reducible $\ul 7 \oplus \ul 1$ representation
of $\qi g_0 = \qthree g_2$, is the irreducible $\ul 8$ of
$\qi g_0' = \q a_2$. Unfortunately, since
\be
  \ul 8 \otimes \ul 8
= \ul 1 \oplus 2 (\,\ul 8\, ) \oplus \ul {10}
  \oplus \ul{\ol{10}} \oplus \ul{27}
\ee
contains two copies of $\ul{8}$, we are not able to construct
$R_{\ul 8, \ul 8}$.

\subsection{Comments}

We have already observed that each $\q g^{(k)}$ multiplet is also a
$\q g^{(1)}$ multiplet, and of course its decomposition does not depend
on whether we choose $g_0$ or $g_0'$. We also observe (as we saw
in the example of $R_{11}$ for $\q a_{2n-1}^{(2)}$ above) that the
$R$-matrices obtained by using $g_0$ and $g_0'$ are similar.

In particular, the $R$-matrices obtained by using $g_0$ and $g_0'$
have the same rank for all values of $x$ and $x'$ respectively, where
$x$ and $x'$ are the spectral parameters in the two versions of the
$R$-matrix and are related by $x^{n_0} = x^{n_0'}$, where the
$n_0,n_0'$ are the Kac marks of the extra root in the two cases,
i.e.\ $x = x'$ except for $g^{(k)}=a_{2n}^{(2)}$.

Heuristically this is most easily seen by recalling that, in
theories with a $\q \hat g$ charge algebra, all the charges
corresponding to step operators transform non-trivially under a
Lorentz boost: the algebra is in the `spin' gradation, equal for
simply-laced algebras to the principal gradation. The similarity
transformation to the homogeneous gradation $R$-matrix then shifts all
the rapidity (spectral parameter) dependence to the generator
corresponding to the extending root. This may be chosen in different
ways, but these will always be similar, and the rank of the direct
channel process cannot depend on this choice.

We also observe that singularities of the $\q g^{(k)}$ $R$-matrix are
also singularities of the $\q g^{(1)}$ $R$-matrix. For example, compare
the $\q e^{(2)}_6$ $R$-matrix above with the $\q e^{(1)}_6$ $R$-matrix for
the  $\ul{27}$ of $\q e_6$ given by the TPG
\[
V_{2\lambda_1}\; \to^1 \; V_2 \; \to^2 \; V_1
\;.
\]
As we see, however, the reverse implication is not true: the $\q g^{(k)}$
$R$-matrix has further singularities, all on links between modules
with the same
parity. This, pointed out in \cite{deliu95b} as the fact that linked
$\qi g_0$ modules have the same parity if they belong to the same
$\qi g$-module, and opposite parities if they belong to different
$\qi g$-modules, allows us to observe that the rational limit
\hbox{$ ( q \to 1 \hbox{ with } x = q^u )$}
of a $\q g^{(k)}$ $R$-matrix is the corresponding $Y(g)$ (Yangian)
$R$-matrix, for all $k$, and not merely, as is well-known, for $k = 1.$

A further observation is that
\be
  g_0^{\vee(1)\vee} = g^{(k)}
\;,
\label{eq:obs}
\ee
for some $k > 1$. (For example, consider $g_0=c_n$ and $g=a_{2n-1}$.)
The significance of this will be drawn out when we
come to discuss affine Toda solitons.

Finally we note that we have not considered all possible
$\qi h \subset \q g^{(k)}$.
For $k=1$, any finite dimensional Lie algebra which is a subalgebra of
$g^{(1)}$ is also a subalgebra of $g$, so that we do not expect to be
able to produce any new $R$-matrices by choosing different $h$;
we expect that any $\q g^{(1)}$ representations which is irreducible
with respect to $\qi h$ will also be irreducible with respect to
$\q g$.
We have also not considered here any non-simple
$\qi h\subset \q g^{(k)}$, e.g.\ those obtained by deleting an interior
node of the $g^{(k)}$ Dynkin diagram.

\section{%
$S$-matrices for affine Toda solitons and the identification of $x$
and $q$.}
\label{sec:threeA}

We recall from \cite{charge} that the $\hat g$ affine Toda field
theories have $\q \hat g^\vee$ charge algebras, that the
quantum solitons are expected to transform as irreducible $\q \hat
g^\vee$ multiplets, and that the $S$-matrix for the scattering of
solitons in irreducible representations $a,b$ is expected to be of the
form
\be
  S_{a,b}(\theta)
= S_{a,b}^{(0)}(\theta)  \;
  \tau_{21}\; \cR_{a,b}^{(TPG)}(x_{a,b}(\theta),q) \;\tau_{12}^{-1}
\;,
\label{smatrix0}
\ee
where $S^{(0)}$ is a scalar factor, $\tau$ denotes the
transformation from the homogeneous to the spin gradation, and
the value of $x_{a,b}(\theta)$ depends on the particular way we
choose to construct the TPG.

{F}rom the explicit construction of the charge algebra given in
\cite{charge}, we can find the $\theta$ dependence of
$x_{a,b}(\theta)$, as detailed in \cite{gande95b}. Provided we choose
an extending root $e_0$ for which the Kac mark is 1, we find
\be
  x_{a,b}(\theta)
= \xi_{a,b}\,
  \exp\left( \left[
         \frac{4 \pi i }{\beta^2} h - h^\vee
             \right] \theta
      \right)
\;,
\ee
where $\xi_{a,b}$ is an overall constant which can be fixed by
demanding crossing symmetry. In many cases $\xi_{a,b}$ can also be
fixed by demanding crossing symmetry in $S_{1,1}$, and then
constructing the other $S$-matrices by the fusion procedure.

It is also possible to deduce that
\be
  q^{ \alpha_i^\vee \cdot \alpha_j^\vee }
= \exp\left(
  \frac{4 \pi^2 i}{\beta^2}\,( \alpha_i^\vee \cdot \alpha_j^\vee )\,
  \right)
\;,
\ee
for all pairs of dual simple roots $\alpha_i^\vee,
\alpha_j^\vee$.
Given our standard root normalizations, it is possible to find a pair
$\alpha_i^\vee,\alpha_j^\vee$ whose inner product is $(-1)$ for all
$\hat g$ except $\hat g = c_n^{(1)}$, and as a result we find
\be
  q
= \cases{
{}~~ \exp(\, \frac{4 \pi^2 i}{\beta^2}\,) &,~~~ $ \hat g \neq c_n^{(1)} $, \cr
\pm\exp(\, \frac{4 \pi^2 i}{\beta^2}\,) &,~~~ $ \hat g = c_n^{(1)} $. }
\ee

Recall now the ansatz used in \cite{gande95b},
\be\label{xq}
\tilde{x}(\theta) = e^{h\lambda\theta}
\;,\hspace{0.3in}
\tilde{q}=e^{\omega i\pi}\;,
\ee
with
\be
  \lambda
= {4\pi\over\beta^2} - {h^\vee\over h}
\;,\hspace{0.3in}
  \omega
={h\over h'} \left({4\pi\over \beta^2}-t\right)
\;.
\ee
This gave the correct masses for direct channel soliton poles at
principal roots of $\tilde{x}=\tilde{q}^r$ and crossed channel poles at
$\tilde{x}=\tilde{x}(i\pi)\tilde{q}^{-r}$, but did not always agree with
crossing symmetry of the $R$-matrices. This is because in some cases
($a^{(2)}_n$ and $c^{(1)}_n$ ATFTs) the correct $R$-matrix to use in
(\ref{smatrix0}), $\cR_{a,b}(x_{a,b}(\theta),q)$, is not equal to
$\cR_{a,b}(\tilde{x}(\theta),\tilde{q})$.


The final result is that the expected soliton masses and crossing
symmetry are all obtained perfectly for $S$-matrices corresponding to
the $R$-matrices in this paper and in \cite{deliu95b}.
There is thus no longer any barrier to the construction of soliton
$S$-matrices corresponding to known $R$-matrices (all those for
$a_{2n}^{(2)}$ ATFTs and $S_{nn}$ in $c_n^{(1)}$ ATFTs), and
we expect to deal with these in a future paper. For the moment,
we limit ourselves to the construction of $S$-matrices for the
$b_n^{(1)}$ affine Toda solitons, the untwisted nonsimply-laced
case for which we have the most $R$-matrix information.

\section{%
$S$-matrices for $b_n^{(1)}$ affine Toda solitons
and their bound states }
\label{sec:fou}

\subsection{Soliton S-matrix}

For details of affine Toda field theories we refer the reader
to our earlier paper \cite{gande95b} and to a recent review \cite{corri94}.
To construct the $S$-matrices for $b_n^{(1)}$ solitons
we recall that the $b_n^{(1)}$ ATFT has a $\q a_{2n-1}^{(2)}$
charge algebra and thus uses the $R$-matrices constructed in
sub-section 4.1. For clarity we give both $\tilde{x}$, $\tilde{q}$
and $x_{a,b}$, $q$ :
$$
\tilde{x}=(-1)^{a+b}x_{a,b}= e^{2n\lambda\theta}\;,
\hspace{0.3in}\hbox{ in which }\;\;\;
\lambda = {4\pi\over \bt^2} - {2n-1\over 2n}\;,
$$
and
\be\label{xqdef}
\tilde{q}=-q=e^{\om i\pi}\;, \hspace{0.3in}\hbox{ in which }\;\;\;
\om={4\pi\over \beta^2}-1 \;,
\ee
and for later convenience we also introduce
\be
\mu \equiv -i\frac{n\la}{\pi}\theta \;.
\ee

We conjecture the exact S-matrix for the scattering of two elementary
solitons in $b_n^{(1)}$ ATFT for $a+b\leq n$ to be
\be
S_{a,b}(\theta) = F_{a,b}(\mu(\theta)) k_{a,b}(\theta)\:
\tau_{21} \cR_{a,b}^{(TPG)}(x_{a,b},q) \tau_{12}^{-1}
\label{smatrix}
\ee
in which $\cR_{a,b}^{(TPG)}(x_{a,b},q)$ is given by (\ref{specdecom})
(and is in this case
equal to $\cR_{a,b}^{(TPG)}(\tilde{x},\tilde{q})$,
because the TPG labels are all odd/even precisely as $a+b$ is
odd/even\footnote{True also for the $d_{n+1}^{(2)}$
soliton $S$-matrices in \cite{gande95b}.}),
$k_{a,b}$ is the $R$-matrix fusion factor and is the same as
in the $d_{n+1}^{(2)}$ case, (5) of \cite{gande95b}, and
the overall scalar factor $F_{a,b}$ will be determined below. The
S-matrix acts as an intertwiner on the modules of the
${}_qd_n$-representations defined in subsection 4.1:
$$
S_{a,b}(\theta): V_{\mu_a}\otimes V_{\mu_b} \to V_{\mu_b}\otimes
V_{\mu_a} \;.
$$
We will denote the
solitons by $A^{(a)}(\theta)$ in which $\theta$ is the rapidity. The
scattering of two solitons of species $a$ and $b$ is then described by
$S_{a,b}(\theta_{12})$, in which $\theta_{12}$ is their rapidity difference.

We then require some explicit crossing and fusion properties
of the $R$-matrices. First, the Birman-Wenzl-Murakami algebra
manipulations analogous to those of appendix B of \cite{gande95b}
for the ${}_qc_n^{(1)}$ $R$-matrices give in the case of
${}_qa_{2n-1}^{(2)}$ $R$-matrices
$$
\cR_{1,1}^{(TPG)cross}(-q^{2n}x_{1,1}^{-1}) =
\frac{(q^2-x_{1,1})(x_{11}+q^{2n})}{(1-x_{1,1})(q^{2n}+x_{1,1}q^2)}
\cR_{1,1}^{(TPG)}(x_{1,1}) \;.
$$
Using (\ref{xqdef}) we can rewrite this equation in terms of $\theta$
and $\la$
$$
c_{1,1}(i\pi-\theta)\cR_{1,1}^{cross}(x_{1,1}(i\pi-\theta)) =
c_{1,1}(\theta)\cR_{1,1}(x_{1,1}(\theta)) \;,
$$
in which
\be
c_{1,1}(\theta) = \sin(\pi(\mu-\om))\sin(\pi(\mu-n\om+\frac12)) \;.
\ee
The general result is then
$$
c_{a,b}(i\pi-\theta)\cR_{a,b}^{(TPG)cross}(x_{a,b}(i\pi-\theta)) =
c_{a,b}(\theta)\cR_{a,b}^{(TPG)}(x_{a,b}(\theta))
$$
in which
$$
c_{a,b}(\theta) =  \prod_{k=1}^b
\sin(\pi(\mu-\fo(a-b+2k)))
\sin(\pi(\mu-\fo(2n-a-b+2k)+\frac12))\;.
$$
This result will be combined with $k_{a,b}$ in the manner described
in \cite{gande95b} in calculating the soliton-breather $S$-matrices.

The construction of the scalar factor $F_{a,b}(\mu)$ is very similar
to the construction in \cite{gande95b} and will therefore only be
described briefly here. $F_{a,b}(\mu)$ must ensure that the S-matrix
satisfies unitarity
\be
S_{a,b}(\theta)S_{b,a}(-\theta) = I_b \otimes I_a \label{unit}
\ee
and crossing symmetry
\be
S_{a,b}(\theta) = [\si S_{b,a}(i\pi-\theta)]^{t_2}\si \;. \label{crossing}
\ee
Considering the case $a=b=1$ first and writing
$F_{1,1}(\mu)=c_{1,1}(\theta)f_{1,1}(\mu)$, the equations (\ref{unit})
and (\ref{crossing}) lead to the following two conditions on $f_{1,1}$:
\bea
f_{1,1}(-\mu+n\la) &=& f_{1,1}(\mu) \label{iter1} \\
f_{1,1}(\mu)f_{1,1}(-\mu) &=&
c_{1,1}^{-1}(\theta)c_{1,1}^{-1}(-\theta) \;. \label{iter2}
\eea
A solution to these equations is given by (see
\cite{gande95b})
\be
f_{1,1}(\mu) = \prod_{j=1}^{\infty} \frac{f^{(1)}[\mu+2n\la(j-1)]
f^{(1)}[-\mu+2n\la(j-\frac12)]}
{f^{(1)}[\mu+2n\la(j-\frac12)]
f^{(1)}[-\mu+2n\la j]} \label{itersol}
\ee
for any function $f^{(1)}(\mu)$ with
$f^{(1)}(\mu)f^{(1)}(-\mu) = c_{1,1}^{-1}(\theta)c_{1,1}^{-1}(-\theta)$.
Writing the right hand side of the equation (\ref{iter2}) in terms
of Gamma functions we can see that an appropriate starting function
$f^{(1)}(\mu)$ is given by:
\be
f^{(1)}(\mu) = \frac1{\pi^2} \Ga(\mu-\om) \Ga(\mu-n\om+\frac12)
\Ga(1+\mu+\om) \Ga(\mu+n\om+\frac12) \;.
\ee
Inserting this into (\ref{itersol}) we obtain the overall scalar
factor of $S_{1,1}$ in the following form
\bea
F_{1,1}(\mu) &=& \prod_{j=1}^{\infty} \frac{\Ga(\mu+2n\la j-\om)
\Ga(\mu+2n\la j-(2n-1)\om)}{\Ga(-\mu+2n\la j-\om)
\Ga(-\mu+2n\la j-(2n-1)\om)} \nn \\
&&\times \frac{\Ga(\mu+2n\la j-n\om+\frac12) \Ga(\mu+2n\la
j-n\om-\frac12)} {\Ga(-\mu+2n\la j-n\om+\frac12) \Ga(-\mu+2n\la
j-n\om-\frac12)} \nn \\
&&\times
\frac{\Ga(-\mu+2n\la j-(n+1)\om-\frac12) \Ga(-\mu+2n\la
j-(n-1)\om+\frac12)} {\Ga(\mu+2n\la j-(n+1)\om-\frac12)
\Ga(\mu+2n\la j-(n-1)\om+\frac12)} \nn \\
&&\times \frac{\Ga(-\mu+2n\la j-2n\om) \Ga(-\mu+2n\la j)}
{\Ga(\mu+2n\la j-2n\om) \Ga(\mu+2n\la j)} \;. \label{F11}
\eea

Starting with $S_{1,1}$ we examine the pole structure of the proposed
S-matrix in order to obtain the poles at which two solitons of species
$a$ and $b$ fuse together to a soliton of species $a+b$ (if
$a+b\leq n$). We find that this process is possible if the rapidity
difference of the incoming solitons is $\theta=
i\frac{(a+b)\pi}n(1-\frac1{2n\la})$. This corresponds to
$\mu = \frac{a+b}2\om$ and we can therefore write the general scalar
factor $F_{a,b}(\mu)$ in the following form
\be
F_{a,b}(\mu) = \prod_{j=1}^a \prod_{k=1}^b F_{1,1}(\mu +
\fo(2j+2k-a-b-2)) \;. \label{Fab}
\ee
These poles correspond to the $R$-matrix singularities due to
horizontal links in the TPG; the vertical links give the
crossed-channel poles.

{F}rom the knowledge of these fusing poles we are also able to deduce
the exact quantum masses of the elementary solitons by using the usual
mass formula
\be
M^2_{a+b} = M^2_a + M^2_b + 2 M_a M_b \cosh(\mbox{Im}\theta).
\label{massform}
\ee
The quantum soliton masses are then
\be
  M_a
= 4\sqrt{2}\frac{Chm}{\beta^2}
  \sin\left(
  \frac{a\pi}{h}\left(1-\frac1{h\la}\right)
  \right)
\label{qumass}
\ee
which could also be obtained from the classical soliton masses
$M_a^{(class.)} = -4\sqrt{2}\frac{hm}{\beta^2}\sin{\frac{a\pi}h}$ by
shifting the Coxeter number $h$ (here $h=2n$) to the so-called quantum
Coxeter number $H$:
$$ h \to H = h+\frac1{\om} \;. $$
Expanding (\ref{qumass}) in terms of $\beta^2$ we obtain
\[
  M_a
= 8\sqrt{2}\frac{Chm}
  {\beta^2}\sin\left(\frac{a\pi}{h}\right)
  [1-\beta^2\frac{a}{4h^2} \cot\left(\frac{a\pi}{h}\right)]
+ O(\beta^2)
\]
which implies that (with an appropriate choice of the overall scale
factor $C$) the mass correction ratios are those of the particles
(l.h.s. of \cite{macka94}, table 5) and thus as suggested in the
note to that paper and not as calculated there for the solitons
(r.h.s. of table 5). We have not yet been able to resolve this
discrepancy, and the semiclassical mass corrections can not therefore
be regarded as understood.

\subsection{Breather S-matrices}

The so-called breathers are scalar bound states of two elementary
solitons%
\footnote{Scalar bound states necessarily have zero topological
charge, but the opposite is not true, as many of the soliton particle
multiplets contain particles of zero topological charge}.
The rapidity difference at which two incoming solitons can fuse
together to a breather can be determined from the spectral
decomposition of the R-matrix (\ref{specdecom}). Since the breathers
transform under the singlet representation the poles are contained in
the prefactor of $\cP_0$, which can only appear in (\ref{specdecom})
if $a=b$. This means that only solitons of the same species can build
scalar bound  states. We will denote the scalar bound state of two
solitons of species $a$ as $B^{(a)}_p$, in which $p$ is an excitation
number.

The construction of the soliton-breather S-matrices and the
breather-breather S-matrices uses the `bootstrap' or
`fusion' procedure and is completely analogous to the construction in
the case of $d_{n+1}^{(2)}$ ATFT as done in \cite{gande95b}. However
in the case of $b_n^{(1)}$ there are some differences regarding the
$n$th breathers. Since we will identify the lowest breather states
with the fundamental quantum particles of the $b_n^{(1)}$ theory in
the next section, we expect the fusion of two solitons of type $n$
into a breather to occur at a different pole than for all other
breathers.

Usually, for algebras where all the $R$-matrices are known (as
applied to the $d_{n+1}^{(2)}$ case \cite{gande95b}), the
fusion poles in the scalar factor mimic the fusion properties
of the $R$-matrix rather beautifully. Here, however, we can only
calculate the $R$-matrices using the TPG for $a+b\leq n$.
Beyond that, pending a general way of treating the higher $R$-matrices,
we only have the information for the scalar factor. In particular our
suggestion for the $n$-th breather pole remains somewhat speculative
and can only be justified by the comparison with the real affine Toda
S-matrices in section 4.3.
We expect the $n$th breather to be roughly half as massive as might
be expected from the na\"{\i}ve $R$-matrix
considerations which led to (46) of \cite{gande95b}. This is not
necessarily a problem, however; for example, it would be resolved
if (the unknown) $R_{n,n}$ were to have singularities at both
$\tilde{x}=-\tilde{q}^{2n}$ (from crossing symmetry) and
$\tilde{x}=\tilde{q}^{2n}$ (expected from
fusion).

We only list the results of the fusion calculations here (for details
of the calculation see \cite{gande95,gande95b}).

Two solitons of type $a$ fuse to give a breather of type $a$ at the
poles:
\bea
\mu &=& n\om-p+\frac12 \hspace{1cm} (\mbox{for } a=1,2,...,n-1; \mbox{
and } p = 1,2,...\leq n\om+\frac12) \nn \\
\mu &=& n\om -\frac p2+\frac12 \hspace{0.9cm} (\mbox{for } a=n; \mbox{
and } p=1,2,...\leq 2n\om+1) \label{nn}
\eea

In order to write down all breather S-matrices in a compact form we
define the blocks
\be
\bl y\br \equiv \frac{\sin(\frac{\pi}{2n\la}(\mu+x))}
{\sin(\frac{\pi}{2n\la}(\mu-x))}
\ee
and
\be
\cbl y \cbr \equiv  \bl y\br \bl n\om+\frac12-y \br \bl \om+1-y\br \bl
n\om-\om -\frac12+y\br
\ee
which have the following properties:
\bea
\cbl n\om+\frac12-y \cbr&=& \cbl y\cbr \hspace{0.5cm} (\mbox{crossing
symmetry}) \nn \\
\cbl y \pm 2n\om\pm1 \cbr &=& \cbl y\cbr \hspace{0.5cm} (2\pi i\mbox{
periodicity}). \nn
\eea

The S-matrix elements for the scattering of breathers are the
following:\\ \\
{\bf Soliton-breather scattering} ($a=1,2,...,n$; $b=1,2,...,n-1$):
\bea
S_{A^{(a)}B^{(b)}_p}(\theta) &=& \prod_{l=1}^p \prod_{k=1}^b \cbl
\fo(2k-a-b+n)+l-\frac p2+\frac14 \cbr \;, \nn \\
S_{A^{(a)}B^{(n)}_p}(\theta) &=&  \prod_{l=1}^p \prod_{k=1}^a
\bl\fo(2k-2-a+2n)+\frac l2-\frac p4-\frac14\br \bl\fo(2k-a)+\frac
l2-\frac p4\br \;. \nn \\ \label{solbrscat}
\eea
{\bf Breather-breather scattering} ($a,b=1,2,...,n-1$):
\bea
S_{B^{(a)}_rB^{(b)}_p}(\theta) &=& \prod_{l=1}^p \prod_{k=1}^b
\cbl \fo(2k-a-b+2n)+l-\frac{p+r}2+\frac12 \cbr  \cbl
\fo(2k-a-b)-l+\frac{p+r}2+1 \cbr \;, \nn \\
S_{B^{(a)}_rB^{(n)}_p}(\theta) &=&  \prod_{l=1}^p \prod_{k=1}^a
\cbl \fo(2k+n-a) +\frac l2-\frac r2-\frac p4 +\frac12\cbr \;,
\nn \\
S_{B^{(n)}_rB^{(n)}_p}(\theta) &=& \prod_{l=1}^p \prod_{k=1}^n \cbl \om k
+\frac l2 -\frac{p+r}4+\frac12 \cbr \;. \label{brbrscat}
\eea

\subsection{Breather-particle identification}

In this section we show that the S-matrix elements
(\ref{brbrscat}) for the lowest breathers ($p=1$), i.e. the breathers
with
lowest mass, coincide with the S-matrices for the fundamental
quantum particles as found for real coupling $b_n^{(1)}$ ATFT by
Delius et al. in \cite{deliu92}. This identification of the lowest
breathers with the
fundamental particles has been demonstrated previously only for the
cases of sine-Gordon theory \cite{zamol79}, $a_2^{(1)}$
\cite{gande95}, $a_2^{(2)}$ \cite{smirn91} and $d_{n+1}^{(2)}$ ATFT
\cite{gande95b}. The identification of the $b_n^{(1)}$ breathers with
the $b_n^{(1)}$ particles shows in particular that there is no
breather-particle Lie duality as suggested in \cite{gande95b}, since
in this latter case the $b_n^{(1)}$ breathers would have to be
identified with the $c_n^{(1)}$ particles instead.

We want to compare the formulas (\ref{brbrscat}) with the S-matrix for
the real $b_n^{(1)}$ ATFT (see \cite{deliu92}):
\bea
S^{(r)}_{ab}(\theta) &=& \prod_{k=1}^b \cbl2k+a-b-1\cbr_H
\cbl H-2k-a+b+1\cbr_H \nn \\
S^{(r)}_{an}(\theta) &=& \prod_{k=1}^a \cbl \frac12 H+2k-a-1\cbr_H \nn \\
S^{(r)}_{nn}(\theta) &=& \frac{-1}{\bl \frac12 B\br_H\bl H-\frac12 B\br_H}
\prod_{k=1}^{n-1} \frac{\bl 2k\br_H\bl H-2k\br_H} {\bl 2k-B\br_H\bl
H-2k+B\br_H}
\label{realsmatrix}
\eea
in which
$$
\cbl y\cbr_H = \frac{\bigl( y-1 \bigr)_H \bigl(
y+1\bigr)_H} {\bigl( y-1+B\bigr)_H
\bigl( y+1-B\bigr)_H} ,\hspace{1.3cm} \bigl(
y\bigr)_H = \frac{\sin(\frac{\theta}{2i}+
\frac{y\pi}{2H})}{\sin(\frac{\theta}{2i}- \frac{y\pi}{2H})}
$$
and $H=h-\frac12 B$, $B=\frac{2\beta^2}{4\pi+\beta^2}$,
$a,b=1,2,...,n-1$.
If we analytically continue $\beta \to i\beta$ we have to change
\beaa
H \to 2n\frac{\la}{\om},&& \hspace{1cm} B\to -\frac2{\om}, \\
\bl y\br_H &\to& \bl \fo y\br\;.
\eeaa
Doing this in the formulas (\ref{realsmatrix}) we obtain as expected
\beaa
S^{(r)}_{ab}(\theta) &\to& S_{B^{(a)}_1B^{(b)}_1}(\theta) \\
S^{(r)}_{an}(\theta) &\to& S_{B^{(a)}_1B^{(n)}_1}(\theta) \\
S^{(r)}_{nn}(\theta) &\to& S_{B^{(n)}_1B^{(n)}_1}(\theta) \;,
\eeaa
and thus establish the identification of the $b_n^{(1)}$ lowest
breathers with the $b_n^{(1)}$ quantum particles.

\subsection{Particle spectrum and figures}

Besides the breathers there are also bound states which non-zero
topological charge present in the theory, which we will call excited
solitons. These bound states correspond to the poles $\mu=a\om-p$ (for
$p=0,1,...\leq a\om$, if $a\leq \frac n2$) at which the S-matrix
element $S_{a,a}(\theta)$
projects onto the representation space $V_{\mu_{2a}}$ (i.e. $S_{a,a}
\sim \cP_{\mu_{2a}}$).

We therefore conjecture that $b_n^{(1)}$ quantum affine Toda field
theory with
imaginary coupling constant contains the following spectrum of
solitons and bound states: \vspace{0.4cm}\\
1) {\em fundamental solitons} $A^{(a)}$ $(a=1,2,..,n)$: \\
 \hspace*{1cm} masses $M_a = C8\sqrt{2}\frac{2nm}{\beta^2}
\sin(\frac{a\pi}{2n}(\frac12-\frac1{2n\la}))$ \vspace{0.2cm} \\
2) {\em breathers} $B^{(a)}_p$ ($A^{(a)}-A^{(a)}$ bound states): \\
 \hspace*{1cm}masses $m_{B_p^{(a)}} = 2M_a\sin(\frac{p\pi}{2n\la})$
 (for $a=1,2,...,n-1$ and $p = 1,2,...\leq n\om+\frac12$):\\
 \hspace*{1cm}and $m_{B_p^{(n)}} = 2M_n\sin(\frac{p\pi}{4n\la})$
 (for $p = 1,2,...\leq 2\om+1$):\vspace{0.2cm} \\
3) {\em excited solitons} $A^{(2a)}_p$ ($A^{(a)}-A^{(a)}$ bound
states) ($p = 0,1,2,...\leq a\om$)\\
\hspace*{1cm} masses $m_{A^{(2a)}_p} =
2M_a\cos(\frac{a\pi}{2n}(1-\frac1{2n\la}-\frac{p}{a\la}))$.

\vspace{.5cm}
We are confident of this conjecture, despite the fact that we can only
check the pole structure of the R-matrices for $a+b \leq n$.
There are a large number of poles which we do not expect to correspond
to the fusion into bound states. As we saw in the $d_{n+1}^{(2)}$
case, it is not enough merely that
simple poles, corresponding to $R$-matrix
singularities, exist. Such poles can often be explained not
as bootstrap poles but in terms of the existing spectrum using
generalized Coleman-Thun methods, and is why odd-species excited
solitons do not seem to occur. Our conjectures can only be verified
by a large body of evidence matching poles to diagrams, and this will
involve much additional work.

The following diagrams show three point vertices involving elementary
solitons, breathers and excited solitons. We were not able to find any
other three
point couplings involving elementary solitons and breathers
only. There are however other vertices involving excited solitons.
Note in particular the differences to the list of three point
couplings in $d_{n+1}^{(2)}$ ATFT \cite{gande95b}: the figures 1c and
1e reflect the slightly different role of the $n$th breather, which
occurs at a different pole from all other breathers. Figure 1g is a
new fusion process which does not exist in the $d_{n+1}^{(2)}$ case. This
process is also consistent with the breather-particle identification,
since the $n$th particle in real $b_n^{(1)}$
ATFT couples to all other particles \cite{brade90}. A generalization of
figure 1g to higher breathers ($p>1$) does not seem to exist.

In the diagrams we have used the abbreviation
$\tm=\mu(i\pi)=n\om+\frac12$.  The imaginary angles correspond to the
rapidity difference of the incoming particles.
Time is meant to be moving upwards in all diagrams, but the processes
obtained by turning any of the diagrams by $120$ degrees are also
allowed. In figure 1a and 1e $a,b$ can take values in $1,2,...,n$
whereas in all other diagrams they take values in $1,2,...,n-1$.

\vspace{1cm}
%
%
%
\begin{center}
\begin{picture}(420,120)(-10,-30)
%

%
\put(0,0){\line(1,1){40}}
\put(40,40){\line(1,-1){40}}
\put(40,40){\line(0,1){64}}
\put(40,40){\circle{16}}
\put(40,17){\vector(0,1){12}}
\put(22,58){\vector(1,-1){10}}
\put(58,58){\vector(-1,-1){10}}
\put(31,7){\shs{\scs{$\frac{a+b}2\om$}}}
\put(0,64){\shs{\scs{$\tm-\frac{b}2\om$}}}
\put(53,64){\shs{\scs{$\tm-\frac{a}2\om$}}}
\put(-5,-10){\shs{\fns{$A^{(a)}$}}}
\put(76,-10){\shs{\fns{$A^{(b)}$}}}
\put(35,108){\shs{\fns{$A^{(a+b)}$}}}
%
%
%
\put(160,0){\line(1,1){40}}
\put(200,40){\line(1,-1){40}}
\put(200,40){\line(0,1){64}}
\put(200,40){\circle{16}}
\put(200,17){\vector(0,1){12}}
\put(182,58){\vector(1,-1){10}}
\put(218,58){\vector(-1,-1){10}}
\put(190,7){\shs{\scs{$\tm-p$}}}
\put(161,64){\shs{\scs{$\frac{\tm}2+\frac{p}2$}}}
\put(214,64){\shs{\scs{$\frac{\tm}2+\frac{p}2$}}}
\put(155,-10){\shs{\fns{$A^{(a)}$}}}
\put(236,-10){\shs{\fns{$A^{(a)}$}}}
\put(195,108){\shs{\fns{$B_p^{(a)}$}}}
%
%
\put(320,0){\line(1,1){40}}
\put(360,40){\line(1,-1){40}}
\put(360,40){\line(0,1){64}}
\put(360,40){\circle{16}}
\put(360,17){\vector(0,1){12}}
\put(342,58){\vector(1,-1){10}}
\put(378,58){\vector(-1,-1){10}}
\put(350,7){\shs{\scs{$\tm-\frac p2$}}}
\put(317,64){\shs{\scs{$\frac{\tm}2+\frac p4$}}}
\put(374,64){\shs{\scs{$\frac{\tm}2+\frac p4$}}}
\put(315,-10){\shs{\fns{$A^{(n)}$}}}
\put(396,-10){\shs{\fns{$A^{(n)}$}}}
\put(355,108){\shs{\fns{$B^{(n)}_p$}}}
\put(17,-30){\shs {\em Figure 1a}}
\put(177,-30){\shs {\em Figure 1b}}
\put(337,-30){\shs {\em Figure 1c}}
\end{picture}

\vspace{1.5cm}
\begin{picture}(420,120)(-10,-30)
%
%
\put(0,0){\line(1,1){40}}
\put(40,40){\line(1,-1){40}}
\put(40,40){\line(0,1){64}}
\put(40,40){\circle{16}}
\put(40,17){\vector(0,1){12}}
\put(22,58){\vector(1,-1){10}}
\put(58,58){\vector(-1,-1){10}}
\put(33,5){\shs{\scs{$\frac{r+p}2$}}}
\put(2,64){\shs{\scs{$\tm-\frac r2$}}}
\put(56,64){\shs{\scs{$\tm-\frac p2$}}}
\put(-5,-10){\shs{\fns{$B^{(a)}_p$}}}
\put(76,-10){\shs{\fns{$B^{(a)}_r$}}}
\put(35,108){\shs{\fns{$B^{(a)}_{p+r}$}}}
%
%
%
\put(160,0){\line(1,1){40}}
\put(200,40){\line(1,-1){40}}
\put(200,40){\line(0,1){64}}
\put(200,40){\circle{16}}
\put(200,17){\vector(0,1){12}}
\put(182,58){\vector(1,-1){10}}
\put(218,58){\vector(-1,-1){10}}
\put(192,5){\shs{\scs{$\frac{p+r}4$}}}
\put(161,64){\shs{\scs{$\tm-\frac r4$}}}
\put(213,64){\shs{\scs{$\tm-\frac p4$}}}
\put(155,-10){\shs{\fns{$B^{(n)}_p$}}}
\put(236,-10){\shs{\fns{$B^{(n)}_r$}}}
\put(195,108){\shs{\fns{$B^{(n)}_{p+r}$}}}
%
%
\put(320,0){\line(1,1){40}}
\put(360,40){\line(1,-1){40}}
\put(360,40){\line(0,1){64}}
\put(360,40){\circle{16}}
\put(360,17){\vector(0,1){12}}
\put(342,58){\vector(1,-1){10}}
\put(378,58){\vector(-1,-1){10}}
\put(352,7){\shs{\scs{$\frac{a+b}2\om$}}}
\put(315,64){\shs{\scs{$\frac{\tm}2-\frac a2\om$}}}
\put(372,64){\shs{\scs{$\frac{\tm}2-\frac b2\om$}}}
\put(315,-10){\shs{\fns{$B_p^{(a)}$}}}
\put(396,-10){\shs{\fns{$B_p^{(b)}$}}}
\put(355,108){\shs{\fns{$B_{p}^{(a+b)}$}}}
\put(17,-30){\shs {\em Figure 1d}}
\put(177,-30){\shs {\em Figure 1e}}
\put(337,-30){\shs {\em Figure 1f}}
\end{picture}

\vspace{1.5cm}
\begin{picture}(420,120)(-10,-30)
%
%
\put(0,0){\line(1,1){40}}
\put(40,40){\line(1,-1){40}}
\put(40,40){\line(0,1){64}}
\put(40,40){\circle{16}}
\put(40,17){\vector(0,1){12}}
\put(22,58){\vector(1,-1){10}}
\put(58,58){\vector(-1,-1){10}}
\put(29,5){\shs{\scs{$\tm-a\om$}}}
\put(0,64){\shs{\scs{$\frac{\tm}2+\frac a2\om$}}}
\put(53,64){\shs{\scs{$\frac{\tm}2+\frac a2\om$}}}
\put(-5,-10){\shs{\fns{$B^{(n)}_1$}}}
\put(76,-10){\shs{\fns{$B^{(n)}_1$}}}
\put(35,108){\shs{\fns{$B^{(a)}_1$}}}
\put(17,-30){\shs {\em Figure 1g}}
%
%
\put(160,0){\line(1,1){40}}
\put(200,40){\line(1,-1){40}}
\put(200,40){\line(0,1){64}}
\put(200,40){\circle{16}}
\put(200,17){\vector(0,1){12}}
\put(182,58){\vector(1,-1){10}}
\put(218,58){\vector(-1,-1){10}}
\put(188,7){\shs{\scs{$a\om-p$}}}
\put(142,64){\shs{\scs{$\tm-\frac{a}2\om+\frac{p}2$}}}
\put(209,64){\shs{\scs{$\tm-\frac{a}2\om+\frac{p}2$}}}
\put(155,-10){\shs{\fns{$A^{(a)}$}}}
\put(236,-10){\shs{\fns{$A^{(a)}$}}}
\put(195,108){\shs{\fns{$A^{(2a)}_p$}}}
\put(177,-30){\shs{\em Figure 1h}}
\end{picture}
\end{center}

\subsection{Comments}

As usual we have little positive evidence of any duality at this
stage. In terms of the duality scheme introduced in \cite{gande95b},
we make the following comments:

\begin{enumerate}

\item[1.]

The $g^{(1)}$ and $g^{\vee(1)\vee}$
quantum soliton multiplets are both $\q g^\vee$ modules,
but they will be different,
e.g.\ the $d_{n+1}^{(2)}$ and $b_n^{(1)}$ soliton multiplets both form
$\q c_n$ modules, but for $d_{n+1}^{(2)}$ these are irreducible,
whereas for $b_n^{(1)}$ they are reducible.

However, as noted earlier, the $\qtwo c_n \subset \qtwo c_n^{(1)}$ and
$\q d_n \subset \q a_{2n-1}^{(2)}$ $R$-matrices used for constructing
their $S$-matrices are related by $q^{2n+2}\mapsto -q^{2n}$.
The basic (vector) representation has the same
dimension in both cases, but the decompositions of its tensor products,
and thus the dimensions of the higher multiplets, are different.
Whether this transformation relates two physical theories is not yet
clear.

\item[2.]

The lowest $b_n$ breathers are to be identified with $b^{(1)}_n$ ATFT
particles, not $c^{(1)}_n$ particles as was implied in
\cite{gande95b}. There is therefore no breather-particle Lie duality. The
particles exhibit a strong-weak affine duality \cite{brade90}, and we now
expect the identification in each ATFT of the particles with the
lowest breathers, which form a small part of a rich spectrum. The
issue thus becomes how this affine duality fits into the overall
picture.

\item[3.]

Application of the $\beta\mapsto 4\pi/\beta$ transformation to higher
breather $S$-matrices does not produce any simple effect.

\end{enumerate}

There may be some significance in (\ref{eq:obs}). (For example,
consider $g_0$ and $g$ to be $c_n$ and $a_{2n-1}$ respectively.)
We obtain the $g_0^{\vee(1)}$ ({\em e.g.\ }$b_n^{(1)}$) ATFT by
folding the $g'^{(1)}$ ({\em e.g.\ }$d_{n+1}^{(1)}$) ATFT, where
$g'^{(1)}$ is the simply-laced algebra corresponding to
$g_0'=g_0^\vee$. Classically the solitons of the former are
multi-solitons of the latter. But the $g_0^{\vee(1)}$ ATFT has a
$\q g_0^{\vee(1)\vee}$ ({\em e.g.\ }$\q a_{2n-1}^{(2)}$) charge
algebra, so that by (\ref{eq:obs}) its soliton multiplets are
multiplets of $\q g^{(k)}$ and therefore, as observed earlier, of $\q
g^{(1)}$ ({\em e.g.\ }$\q a_{2n-1}^{(1)}$). Should they be identified
with a subset of
the single $g^{(1)}$ solitons? They have the same classical masses and
quantum multiplets.  One could not expect the quantum masses to be the
same, since the corrections to the latter would now be calculated
using only a subspace of the full $g^{(1)}$ theory. But there are also
more fusings than would be expected with this identification: to give
another example, in the $\q e^{(2)}_6$ case discussed above, the
$e^{(1)}_6$ theory had  $11 \to 1$ and $11 \to 3$ fusings, but the
$\q e^{(2)}_6$ $R$-matrix and thus the $f_4^{(1)}$ soliton had in
addition a $11 \to 2$ fusing. We leave this as a tantalising fact
to be enlarged upon.

\vspace{0.1in}
{\bf Acknowledgments}

\noindent We would like to thank M.Jimbo and R.Nepomechie for their
comments.
NJM would like to thank PPARC (UK) for a research fellowship,
NSERC (Canada) for an exchange fellowship and the CRM for its
hospitality during the course of this work.
GMTW was supported by a NSERC (Canada) international fellowship.

\end{document}